\documentclass[aps,prd,twocolumn,showpacs,superscriptaddress,groupedaddress]{revtex4}  
\usepackage{graphicx}  
\usepackage{dcolumn}   
\usepackage{bm}        
\usepackage{amssymb}   
\usepackage{amsmath}   

\begin{document}



\title{Einstein-Yang-Mills-Dirac systems from the discretized Kaluza-Klein theory}
\author{Nguyen Ai Viet} \affiliation{ITI, Vietnam National University, Hanoi, Vietnam} \affiliation{Physics Department, College of Natural Sciences, Vietnam National University, Hanoi, Vietnam}
\author{Nguyen Van Dat} \affiliation{ITI, Vietnam National University, Hanoi, Vietnam} \affiliation{Physics Department, College of Natural Sciences, Vietnam National University, Hanoi, Vietnam}
\author{Nguyen Suan Han} \affiliation{Physics Department, College of Natural Sciences, Vietnam National University, Hanoi, Vietnam}
\author{Kameshwar C. Wali} \affiliation{Physics Department, Syracuse University, Syracuse, New York, USA}

\date{\today}

\begin{abstract}
A unified theory of the non-Abelian gauge interactions with gravity in the framework of a discretized Kaluaza-Kleine theory is constructed with a modified Dirac operator and wedge product. All the couplings of chiral spinors to the non-Abelian gauge fields emerge naturally as components of the couplings of the chiral spinors to the generalized gravity together with some new interactions. In particular, the currently prevalent gravity-QCD-quark and gravity-electroweak-quark-lepton models are shown to follow as special cases of the general framework.   

\end{abstract}

\maketitle


\section{Introduction}
The discretized Kaluza-Klein theories(DKKT) have provided a new framework to yield a non-supersymmetric, perturbatively stable Standard Model at the $TeV$ scale. The gauge theory of the electroweak interactions, where the symmetry is broken spontaneously with a quartic Higgs field potential, is shown to follow from a dimensional destruction by discretizing the extra internal dimensions \cite{ACG2001a,ACG2001b}. The framework and techniques can also be extended to construct an effective multi-gravity theory with finite massive gravitons up to the Planck scale \cite{AGS2003,DM2005}. Recently, the ghost-free models of massive gravity and their multigravity extensions follow from considering DKKT \cite{dRMT}

In fact, DKKT had been proposed sometime ago \cite{LVW1994} and developed further in \cite{VW1995a, VW1995b}, utilizing the mathematical concepts of noncommutative geometry (NCG) {\it \`a la Connes} \cite{Connes}. In this framework the compact fifth dimension of the conventional Kaluza-Klein theory \cite{KK} is replaced by two discrete points. The generalized Cartan-Einstein-Hilbert action leads to the same zero mode sector of the ordinary Kaluza-Klein theory consisting of gravity, electromagnetism and a Brans-Dicke scalar \cite{LVW1994}. The advantage of the discretized version over the ordinary one is that it avoids the existence of the infinite towers of massive fields, which might lead to both theoretical and observational obstacles. In fact, the torsion free version \cite{VW1995a} of this theory contains just two dilatons in addition to the zero mode sector. A more general metric compatible theory \cite{VW1995b} leads to pairs of tensor, vector and scalar fields. In each pair, one field is massless and the other is massive. 

The space-time of DKKT can also be viewed as two parallel copies of the conventional one. Connes and Lott \cite{CoLo} proposed a  two sheeted space-time of ${\cal M}^4 \times Z_2$ as a realistic approach to formulate  the Standard Model of electroweak and strong interactions of chiral quarks and leptons. In their approach , based on NCG, the gauge fields and the scalar Higgs fields appear on the same footing. The requires quartic Higgs potential emerges naturally to trigger the spontaneous symmetry breaking. It has many new features leading to the predictions of the parameters of the SM such as Higgs and Top quark masses \cite{MRST94} to name just a few. These and the inclusion of gravity with significant new results leads us to pursue the more exciting perspective of a new concept of space-time that can lay the foundation of for all the existing models and pave the way to new physics. 

Along this direction, the couplings of the matter to gravity and gauge fields have been discussed in \cite{VW2000, VW2003}. Since the gauge vectors in the gravity sector are abelian, the nonabelian gauge fields have been introduced into the theory in an independent term. Since the ordinary Kaluza-Klein theory can be generalized to contain the nonabelian gauge vectors, one expect that the same situation is possible in the discretized one. Recently, Viet and Du \cite{VietDu2015} have shown that DKKT can be generalized further to have the nonabelian gauge fields as components of gravity. Interestingly, the generalization is possible only with certain limited gauge symmetries, including those of the strong and weak interactions. In other words, the space-time extended by discrete dimensions can provide explanations for the chiral symmetry of the strong interaction and for the absence of the coupling between the right handed chiral spinors and the nonabelian gauge bosons of the weak interaction. This result can be interpreted as an explanation for a model to unify all the known interactions and Higgs fields as components of gravity in the new space-time of ${\cal M}^4 \times Z_2 \times Z_2$ \cite{QuyNhon2015, Viet2015}. 

Although the perspective of a complete, satisfactory unified theory of all interactions is fascinating, a unified mathematical framework for gravity and other interactions seems to be the necessary first step. This has been the main purpose of our paper. So far the approaches in DKKT theories and the construction of spontaneously broken Yang-Mills theories differ in details in formulating the necessary differential forms and constructing the necessary quartic Higgs potential, sometimes appearing ad hoc. This is avoided in the current framework, providing a consistent framework bringing together the discretized extra dimension lending to non-Abelian gauge fields and the four-dimensional continuum space-time for gravity.   

The paper is organized as follows: In the next section, we set up the basic mathematical framework of NCG that generalizes the conventional differential calculus to DKKT space-time and recovers the results of Connes-Lott’ and other DKKT models. Section III is devoted to the gravity sector and the derivation of the Cartan- Einstein-Hilbert action including non-vanishing torsion terms subject to a set of constraints \cite{VW1995b, Viet2014}.  In Section IV, we discuss the couplings of the chiral spinors (matter fields) to the vector and scalar fields of the generalized gravity in the extended space-time. The final section is devoted to a summary and discussion of the results.          

\section{General framework of the DKKT NCG space-time}

In this paper we will confine ourselves with a discretized fifth extra-dimension consisting of two points, keeping in mind that it is straight forward to generalize it to the case of arbitrary N points. 

\subsection{The spectral triple}

The generalized functions now are a pair of the conventional ones defined at two different internal points.  In both NCG and dimensional deconstruction approaches to DKKT \cite{ACG2001a,ACG2001b,AGS2003,DM2005,dRMT,LVW1994, VW1995a,VW1995b}, the derivation over the discrete dimension is defined as the difference of this pair of functions multiplied by a mass parameter $m$ that carries the inverse dimension of distance between the two points.

The dimensional deconstruction approach defines the Lagrangian based on the intuitive concepts of the site and link interactions, while the NCG formalism is based on the geometric concepts. 

In the general framework of NCG, the concept of space-time is freed from the notions of coordinates and derivations along the coordinates. Instead, it is defined by the spectral triple as its building blocks: a Hilbert space of the wave functions of matter fields, an algebra of fields, an algebra of functions that act as operators on the Hilbert space and an operator called as Dirac operator  that encodes all the information about the space-time manifold. In LVW approach to NCG, the concept of derivative over the discrete dimension has been included in the Dirac operator to construct the geometric motions in a more transparent analogy with the conventional Riemannian geometry. 

In this paper, we will adopt the Connes-Lott's two sheeted space time \cite{CoLo} of left and right handed chiral spinors generalized to our two arbitrary sheets \footnote{Actually, in the DKKT which unifies of all interactions proposed recently, one additional discrete dimension leading to new type of spinor has been used \cite{QuyNhon2015, Viet2015}}.

i) The Hilbert space $ {\cal H} = {\cal H}_L \oplus {\cal H}_R$. Thus, the generalized spinors in our model are direct sums of two chiral spinors of the 4-dimensions, 
\begin{equation}\label{spinor}
\Psi =
\begin{bmatrix}
    & \psi_L & \\
    & \psi_R &
\end{bmatrix}
\end{equation}

ii) The algebra of the function operators $ {\cal A} = {\cal A}_L \oplus {\cal A}_R$, where ${\cal A}_I= {\cal C}^\infty({\cal M}^4), I=L,R$, whose elements can be represented as the diagonal 0-form matrices ${\cal F}$, with ${\cal F}$ being elements of ${\cal A}$, are real functions, where  
 
\begin{equation}\label{Function}
F(x) = 
\begin{bmatrix}
    f_L(x) & 0 \\
    0 & f_R(x)
\end{bmatrix}.
\end{equation} 

iii) The Dirac operator $D = d. {\bf e} + \Theta $ is an extension of the usual Dirac operator $d = \gamma^\mu \partial_\mu$ in the ordinary space-time ${\cal M}^4$, defined by 

\begin{equation}\label{DIRAC2}
D = \begin{bmatrix}
\gamma^\mu \partial_\mu    & -i m \gamma^5 \cr
i m \gamma^5  & \gamma^\mu \partial_\mu 
\end{bmatrix} ~,~ 
\Theta = \begin{bmatrix}
0  & -i m \gamma^5 \cr
i m \gamma^5  & 0
\end{bmatrix}, 
\end{equation}
whose action on the generalized function $F$ leads to
\begin{eqnarray}\label{DERNEW}
DF = [D, F]~~~~~~~~~~~~~~~~~~~~~~~~~
~~~~~~~~~~~~~~~~~~~~~~~&& \nonumber \\
=
\begin{bmatrix}
\gamma^\mu \partial_\mu f_L(x)   &  im \gamma^5 ( f_L(x) - f_R(x))\\
i m \gamma^5 ( f_L(x)- f_R(x))  & \gamma^\mu \partial_\mu f_R(x) 
\end{bmatrix},&& 
\end{eqnarray}
 
The new Dirac operator leads to the desired Lagrangian of the massive and massless chiral spinors as follows
\begin{widetext}
\begin{eqnarray} \label{goodlag}
{\cal L}_\Psi &=&  i {\bar \Psi} D \Psi = i {\bar \psi}_L \gamma^\mu \partial_\mu \psi_L + i {\bar \psi}_R \gamma^\mu \partial_\mu \psi_R +  m {\bar \psi}_L \psi_R + m {\bar \psi}_R \psi_L 
=  {\bar \psi}( i\gamma^\mu \partial_\mu + m)\psi  ,
\end{eqnarray}
\end{widetext}

\subsection{The generalized 1-forms}

Following the conventional geometry, we rewrite the exterior derivative of 0-forms in Eq.(\ref{DERNEW}) in the form
\begin{eqnarray}\label{eq:DF}
DF & = & DX^\mu [D_\mu, F(x)] + DX^5 \sigma^\dagger [D_5, F] \nonumber \\
&=& DX^\mu \partial_\mu F(x) + DX^5 \sigma^\dagger \partial_5 F  \label{D0Form}, 
\end{eqnarray}

where
\begin{equation}
D_5 = \begin{bmatrix}
0 & -m \cr
m & 0 
\end{bmatrix} ~,~ 
\sigma^\dagger = \begin{bmatrix}
0 & -1 \cr
1 & 0 
\end{bmatrix}.  
\end{equation}

In the Dirac matrix representation, the differential elements $DX^\mu$ are represented by the following generalized $\gamma$-matrices
\begin{equation}
\Gamma^\mu = \begin{bmatrix}
\gamma^\mu & 0 \cr
0 & \gamma^\mu 
\end{bmatrix} ~,~ \Gamma^5 = \begin{bmatrix}
0 & i\gamma^5 \cr
-i \gamma^5 & 0
\end{bmatrix}.
\end{equation}
 
The general 1-forms $U$ then take the form 
 \begin{equation}\label{H1FORM}
 U = \Gamma^\mu U_\mu + \Gamma^{{5}} U_{{5}} = \begin{bmatrix}
 \gamma^\mu u_{\mu L} & i\gamma^5 u_{5R} \cr
 -i\gamma^5 u_{5L} & \gamma^\mu u_{\mu R},
 \end{bmatrix} 
 \end{equation}
where $U_M$ are generalized functions. So, the 1-form $U$ contains two vectors and two scalar functions.

The exterior derivative of a function given in Eq.(\ref{DERNEW}) is also a 1-form with a special condition  $u_{5L} = - u_{5R}$. 

For incorporating gravity, we are in particular interested in hermitian 1-forms $U$, with the condition $u_{5L}= u_{5R}$. Obviously, in our framework the hermitian 1-forms automatically are not derivatives of a 0-form.

The fifth derivative in Eq.(\ref{eq:DF}) has been used in the "link" interaction terms of the dimensional deconstruction approach \cite{ACG2001a, ACG2001b, dRMT} as an ad hoc derivative over the discrete dimension. 

\subsection{Wedge products and 2-forms}

To construct 2-forms from the derivative of 1-forms, one adopts generally in the continuum 4-D space, the completely anti-symmetric wedge product to avoid the so called “junk-forms” (for details see \cite{Connes}). In our case, with the extra discretized internal space, we have a choice that leads us to the following definitions for the wedge products in Eq.(\ref{Wedge2}). As we shall see, in addition to the gauge and scalar fields being on the same footings,  it leads  naturally to a quartic Higgs potential that is required for spontaneous symmetry breaking.

\begin{eqnarray}\label{Wedge2}
 DX^\mu {\wedge} DX^\nu & =& - DX^\nu {\wedge} DX^\mu \nonumber \\
 DX^\mu {\wedge} DX^5 & = & - DX^5 {\wedge} DX^\mu  \nonumber \\
 DX^5 {\wedge} DX^5 & \not = & 0,
 \end{eqnarray}
\begin{eqnarray}\label{Wedge1forms}
(U \wedge V)_{\mu \nu} &=& - (U \wedge V)_{\nu \mu} = {1 \over 2}(U_\mu V_\nu - U_\nu V_\mu) \nonumber \\
(U \wedge V)_{\mu 5} &=& - (U \wedge V)_{5 \mu} = {1 \over 2}({\tilde U}_\mu V_5 - U_5 V_\mu) \nonumber \\
(U \wedge V)_{55}& =& {\tilde U}_5 V_5 
\end{eqnarray}
where the "\~{}" operation is defined as
\begin{eqnarray}
{\tilde F} &=& f_+ {\bf e} - f_-{\bf r} ~,~ f_\pm = {1 \over 2}(f_1 \pm f_2) \nonumber \\
{\bf e} &=& \begin{bmatrix}
1 & 0 \cr
0 & 1 \cr
\end{bmatrix}~,~ {\bf r} = \begin{bmatrix}
1 & 0 \cr
0 & -1 \cr
\end{bmatrix}
\end{eqnarray}

The exterior derivative of 1-form is given as follows
\begin{eqnarray}\label{eq:DU}
DU & =& [D,U] = DX^M \wedge DX^N (DU)_{MN} \\
(DU)_{\mu \nu} &=& {1 \over 2}(\partial_\mu U_\nu - \partial_{\nu} U_\mu) = - (DU)_{\nu \mu} \nonumber \\
(DU)_{\mu 5} &=& {1 \over 2}(\partial_\mu U_5 - m( U_\mu - {\tilde U}_\mu)) = - (DU)_{5\mu} \nonumber \\
(DU)_{55} & = & - m(U_{5L}+U_{5R})
\end{eqnarray} 

It is straightforward to verify the defined wedge products lead to the de Rham’s condition $D^2 =0$ to avoid the ‘junk” forms. Also, as said before, it is motivated by the required condition that the Higgs Lagrangian contain a quartic potential.    

\section{Setting up the curved space-time}
\subsection{Vielbein and metric}

Einstein’s general theory of gravity begins with the assumption of a locally flat inertial system built around an orthonormal coordinate system described in terms of vierbein. Consistent with the principle of equivalence, it allows to define the metric that defines the distance between two points and affine connections and extend to curved space time. Vierbein formulation also provides the necessary spin connections to couple fermionic matter fields. We can generalize this concept to NCG by defining a locally flat frame $E^A$ and its curvilinear one $DX^M$.
\begin{eqnarray}\label{Transform}
E^A = DX^M E^A_M(x) &,&
DX^M = E^A E^M_A(x) \nonumber \\
E^A_M(x) E^N_A(x) = \delta^N_M(x) &,& E^A_N(x) E^N_B(x) = \delta^A_B(x) 
\end{eqnarray}

The inner scalar products of 1-forms ${\cal U}$ and ${\cal V}$ are defined accordingly
\begin{eqnarray}\label{Metric}
&<DX^M, DX^N> = G^{MN}(x)& \nonumber \\
&<U, V> =  U^\dagger_M G^{MN}(x) U_N&
\end{eqnarray}
The locally flat inertial coordinate system is chosen to satisfy
\begin{equation}\label{Flatcond}
<E^A,E^B> = \eta^{AB} = diag(-1,1,1,1,1)
\end{equation}
and using the above condition we can express the metric $G^{MN}$ in terms of vielbeins
\begin{equation}\label{Metric}
G^{MN}(x) = E^M_a(x) \eta^{ab} E^N_b(x) + E^M_{\dot{5}}(x) E^N_{\dot{5}}(x)
\end{equation}
In the locally flat frame, with the above bases, we find the Dirac matrix representation as given by
\begin{eqnarray}\label{Vielbein}
DX^\mu & = & \begin{bmatrix}
\gamma^a e^\mu_{a L}(x) & 0 \cr
0 & \gamma^a e^\mu_{a R}(x)
\end{bmatrix}, \nonumber \\
 DX^5 &=& \begin{bmatrix}
-\gamma^a e^\mu_{a L}(x) a_{\mu L}(x) & i\gamma^5 \phi^{-1}(x) \cr
-i \gamma^5 \phi^{-1}(x) & -\gamma^a e^\mu_{a R}(x) a_{\mu R}(x)
\end{bmatrix},\nonumber \\ 
E^a &=& \begin{bmatrix}
\gamma^a & 0 \cr
0 & \gamma^a
\end{bmatrix}
~,~ E^{\dot{5}} = 
\begin{bmatrix}
0 & i \gamma^5 \cr
-i \gamma^5 & 0 
\end{bmatrix}.
\end{eqnarray} 
The transformation coefficients between the locally flat and curvilinear frames encode gravity are the vielbein matrices 
\begin{eqnarray}\label{Vielbein2}
E^a_\mu(x) &=& \begin{bmatrix}
e^a_{\mu L}(x)& 0 \cr
0 & e^a_{\mu R}(x)
\end{bmatrix},~~ E^a_5=0,~~ E^{\dot{5}}_{5} = 
\phi(x), \nonumber  \\
 E^{\dot{5}}_\mu(x) &=& A_\mu(x)\phi=
\begin{bmatrix}
a_{\mu L}(x)& 0 \cr
0 & a_{\mu R}(x) 
\end{bmatrix} \phi, 
\end{eqnarray} 
and their inverses
\begin{eqnarray}\label{Vielbein2inverse}
 E^\mu_a(x) = \begin{bmatrix}
e^\mu_{aL}(x)& 0 \cr
0 & e^\mu_{aR}(x)
\end{bmatrix}
,& E^\mu_{\dot 5}(x) = 0,&  \nonumber \\
 E^5_a(x) = - E^\mu_a(x) A_\mu(x) 
,& E^5_{\dot 5} = \phi^{-1}(x)
& 
\end{eqnarray} 
The generalized vielbein of this theory contains two vierbeins $e^a_{I\mu}(x), I=L,R$, two possibly nonabelian vectors $a_{I \mu}(x)$ and one real Brans-Dicke scalar $\phi(x)$.

In contrast to the case of pure abelian case in Ref.\cite{LVW1994, VW1995a, VW1995b}, here we allow the vierbeins, vectors and scalar to have the matrix values (diagonal and non-diagonal for the possibly non abelian fields). The $Z_2$ structure of the fifth dimension still valid, since it represents the space-time sheets of right and left-handed chiral quark-leptons. Further dimensions leading to the matrix values can be considered as the hyperfine structure of current discretized five space-time as stated in \cite{CoLo, MRST94}.

In the rest of this paper, we will use the following notations
\begin{eqnarray} \label{eq.useful}
h_{a\mu} &=& {1 \over 2} (E_{a\mu} - {\tilde E}_{a\mu}) {\bf r} ~~,~~
e^\mu_a  = {1 \over 2} (E^\mu_a 
+ {\tilde E}^\mu_a) \nonumber\\
g^{\mu \nu} &=& e^\mu_a \eta^{ab} e^\nu_b ~,~ {\hat f}_{I \mu \nu} = \partial_\mu a_{I\nu} - \partial_\nu a_{I \mu}\nonumber \\
f_{I \mu \nu} &=& \partial_\mu a_{I\nu} - \partial_\nu a_{I \mu} + m [a_{I\mu}, a_{I \nu}], I=\pm, L, R  \nonumber \\
F_{\mu \nu} &=& \partial_\mu A_\nu - \partial_\nu A_\mu = {\hat f}_{+\mu \nu} + {\hat f}_{-\mu\nu}{\bf r} \nonumber \\
a_{-\mu} &=& {1 \over 2} (A_\mu - {\tilde A}_\mu) {\bf r} ~~,~~
a_{+\mu} = {1 \over 2} (A_\mu + {\tilde A}_\mu) 
\end{eqnarray}
So that when $E^a_\mu = {\tilde E}^a_\mu$, the gravity sector reduces to the conventional one with the vielbein $e^\mu_a$ and $h_{a\mu}=0$. In the bigravity model, one can choose $h_{a\mu}$ and $e^\mu_a$ to represent two independent fields as an alternative to $e^a_{L\mu}$ and $e^a_{R\mu}$.

\subsection{Levi-Civita connections and Cartan's structure equations}

The generalized Levi-Civita connection 1-forms $\Omega^A_{~B}$ follow from the covariant derivative of the generalized 1-form $E^A$
\begin{equation}
\nabla E^A =E^B \otimes \Omega^A_{~B},
\end{equation}

The metric compatibility and hermitian conditions require
\begin{equation}\label{metriccomp}
\Omega^\dagger_{AB} = \Omega_{AB} = - \Omega_{BA}
\end{equation}

The generalized Cartan's structure equations take the form
\begin{eqnarray}
T^A &=& DE^A - E^B \wedge \Omega^A_{~B} \label{Struct1} \\
R^{AB}&=& D\Omega^{AB} + \Omega^A_{~C} \wedge \Omega^{CB} \label{Struct2}
\end{eqnarray}

Next, we impose, as in Ref. \cite{VW1995b} a set of minimal conditions
\begin{equation}
T_a = 0 ~,~  T_{{\dot 5}ab} = t_{{\dot 5} ab} {\bf r} ~,~ T_{{\dot 5} a{\dot 5} } = t_{{\dot 5} a {\dot 5}} {\bf r},
\end{equation}
to obtain all the Levi-Civita connections in terms of metric vielbeins
\begin{widetext} 
\begin{eqnarray}\label{Omega}
\Omega_{abc} &=& (DE_a)_{bc} +  (DE_b)_{ca}-(DE_c)_{ab}~~,~~
\Omega_{{\dot 5}a {\dot 5}} =  e^\mu_a{\partial_\mu \phi(x) \over \phi} \nonumber \\
\Omega_{{\dot 5}ab} &=& {1 \over 2}E^\mu_a E^\nu_b \phi (  F_{\mu \nu} - 2m ( A_\nu a_{-\mu} - A_\mu a_{-\nu} ){\bf r})+ m \phi^{-1} ( 3 h_{a\mu} e^\mu_b  - h_{b\mu} e^\mu_a) {\bf r} \nonumber \\
&& -
{1 \over 4}\phi
(E^\mu_a E^\nu_b F_{\mu \nu}
- {\tilde E}^\mu_a {\tilde E}^\nu_b{\tilde F}_{\mu \nu} 
 - 2m(E^\mu_a E^\nu_b(A_\nu a_{-\mu} - A_\mu a_{-\nu})
 +  {\tilde E}^\mu_a {\tilde E}^\nu_b ({\tilde A}_\nu a_{-\mu} - {\tilde A}_\mu a_{-\nu})  ){\bf r})
\end{eqnarray}
\end{widetext}
Substituting the above results in Eq.(\ref{Struct2}) we obtain the components of the curvature tensors $R_{AB}$
\begin{widetext}
\begin{eqnarray} \label{RABMN}
R_{abcd} &=&  (D\Omega_{ab})_{cd} + \eta^{ef}(\Omega_{ae} \wedge \Omega_{fb})_{cd} + (\Omega_{a{\dot 5}} \wedge \Omega_{{\dot 5}b})_{cd} 
= {1 \over 2}E^\mu_c E^\nu_d (\partial_\mu( E^e_\nu \Omega_{abe})-\partial_\nu (E^e_\mu  \Omega_{abe})\nonumber \\
&& + \phi (F_{\mu \nu}+ m [A_\mu, A_\nu]) \Omega_{ab{\dot 5}} 
 - m ( (A_\nu  E^e_\mu - A_\mu E^e_\nu) \Omega_{abe} + (A_\mu{\tilde E}^e_\nu - A_\nu{\tilde E}^e_\mu ){\tilde \Omega}_{abe} \nonumber \\
&&  + \phi (A_\mu {\tilde A}_\nu - A_\nu {\tilde A}_\mu ){\tilde \Omega}_{ab{\dot 5}})) -
{1\over 2}(\Omega^e_{~ac}  \Omega_{ebd} - \Omega^e_{~ad} \Omega_{ebc}) - {1 \over 2} 
(\Omega_{{\dot 5}ac} \Omega_{{\dot 5}bd} - \Omega_{{\dot 5}ad} \Omega_{{\dot 5}bc})  \nonumber \\
R_{{\dot 5}a{\dot 5}b} & = & (D\Omega_{{\dot 5}a})_{{\dot 5}b} + \eta^{cd}(\Omega_{{\dot 5}c} \wedge \Omega_{da})_{{\dot 5} b} = - {m \over 2}\phi^{-1} e^\mu_b(E^c_\mu\Omega_{{\dot 5}ac}- {\tilde E}^c_\mu {\tilde \Omega}_{{\dot 5}ac}) - {1 \over 2} \eta^{cd} {\tilde \Omega}_{{\dot 5}cb} \Omega_{da{\dot 5}}  \nonumber \\
R_{{\dot 5} {\dot 5} {\dot 5}{\dot 5}} &=& 
\eta^{cd}(\Omega_{{\dot 5} c} \wedge \Omega_{d {\dot 5}})_{{\dot 5}{\dot 5}} = \eta^{cd}{\tilde \Omega}_{c{\dot 5} {\dot 5}} \Omega_{d{\dot 5}{\dot 5}} = g^{\mu \nu} {\partial_\mu \phi \over \phi} {\partial_\nu \phi \over \phi }
\end{eqnarray}
\end{widetext}

\subsection{The Hilbert-Einstein  action}

We begin by defining the $5D$ Ricci curvature and the covariant integration in the $5D$ space-time as follows
\begin{equation}
R_5 =   {1 \over 2} Tr (\eta^{AC} R_{ABCD} \eta^{BD})
\end{equation}

The covariant integration in the 5D space-time is defined as follows
\begin{equation}
\int DX^5 \sqrt{-det G} = \int dx^4 \sqrt{-det g}~ \phi,  
\end{equation}
where $ det g $ is the usual 4D determinant calculated from the average metric on the two sheets $g^{\mu \nu}$ given in Eq.(\ref{eq.useful}). 
 
The generalized Hilbert-Einstein action is defined as follows
\begin{equation}\label{eq:SHE}
S_{HE}(5) = M_{Pl}^2(5) \int dx^4 \sqrt{-det g}~\phi_0 \exp{({\sigma(x) \over m_\sigma})}~ R_5 ,
\end{equation}
where $m_\sigma$ is a mass parameter for giving physical dimension to the scalar field $\sigma(x)$, which is introduced by the field redefinition
\begin{equation}
\phi = \exp{ ({\sigma(x) \over m_\sigma} + \kappa_0)}.
\end{equation}
Hence, the $5D$ Planck mass $M_{Pl}(5)$ is related to the $4D$ one follows.
\begin{equation}
M_{Pl}(5) = M_{Pl}(4) \exp{(-{\kappa_0 \over 2})}
\end{equation}
Hence, $M_{Pl}(5)$ can be many orders lower than the conventional Planck scale $M_{Pl}(4)$. 

The generalized Ricci scalar curvature $R_5$ can be calculated explicitly to yield
\begin{widetext}
\begin{eqnarray}\label{eq:R5}
R_5 &= & {1 \over m^2_\sigma} g^{\mu \nu} \partial_\mu \sigma(x) \partial_\nu \sigma(x) +  {1 \over 2} Tr(
\begin{bmatrix}
r_L & 0 \cr
0 & r_R 
\end{bmatrix} 
 - {1 \over 2} ((\Omega_{{\dot 5}ac} \eta^{ac})^2 - \Omega^{{\dot 5}ab} \Omega_{{\dot 5}ba}) \nonumber \\
&+&
 {1 \over 2}E^{a\mu} E^{b\nu} (
\phi (F_{\mu \nu}+ m [A_\mu, A_\nu]  - m (A_\mu {\tilde A}_\nu - A_\nu {\tilde A}_\mu ) ) \Omega_{ab{\dot 5}}- m ( (A_\nu  E^e_\mu - A_\mu E^e_\nu) \Omega_{abe} \nonumber \\
&+& (A_\mu{\tilde E}^e_\nu - A_\nu{\tilde E}^e_\mu ){\tilde \Omega}_{abe}))) 
+{1 \over 2} Tr(- m \phi^{-1} e^{a\mu}( E^b_\mu \Omega_{{\dot 5} ab } - {\tilde E}^b_\mu {\tilde \Omega}_{{\dot 5} ab }) - {\tilde \Omega}^{{\dot 5}ab} \Omega_{ab{\dot 5}})  
\end{eqnarray}
\end{widetext}

The first term in the above equation represents clearly the kinetic term for the physical field $\sigma(x)$ if one choose the mass parameter $m_\sigma$ as follows
\begin{equation}
m_\sigma = {M_{Pl}(4) \over \sqrt{2}}.
\end{equation}
At this point, we can conclude that
the Hilbert-Einstein action defined in Eq.(\ref{eq:SHE}) contains bigravity, two possibly nonabelian gauge fields and one Brans-Dicke scalar. Since in this paper, we focus mainly on the Einstein-Yang-Mills-Dirac systems, we will discuss the issues related to bigravity elsewhere. The Cartan-Hilbert-Einstein's action with the contribution from the torsion is not considered in this paper, as it violates the gauge invariance.

\subsection{Gravity and nonabelian gauge fields}

Now focusing on single gravity coupled to nonabelian gauge fields and the Brans-Dicke field, we choose the vierbeins
\begin{eqnarray}\label{eq:samegravity}
e^a_{\mu L} &=& e^a_{\mu R} = e^a_\mu(x),
\end{eqnarray}

The curvature  $R_5$ in Eq.(\ref{eq:R5}) now is reduced to the following expression
\begin{eqnarray}\label{TrR5}
R_5& =&   r_4  - {1 \over 16} \phi^2 g^{\mu \rho}  g^{\nu \tau} {\cal F}_{\mu \nu} {\cal F}_{\rho \tau} \nonumber \\
&&
+ {1 \over 2M^2_{Pl}(5)}g^{\mu \nu} \partial_\mu \sigma(x) \partial_\nu \sigma(x) , \\
{\cal F}_{\mu \nu}& =&  f_{L\mu\nu}  +f_{R\mu\nu}  + m ( [a_{R\nu}, a_{L\mu}] + [a_{L\nu}, a_{R\mu}]),~~~~
\end{eqnarray}
where the gauge covariant field strengths are defined as follows
\begin{eqnarray}
f_{I\mu\nu} &=& \partial_\mu a_{I\nu} -\partial_\nu a_{I\mu} + m[a_{I\mu}, a_{I\nu}],~I=L,R~~.
\end{eqnarray}
 
In general, due to the extra terms, the "field strength" ${\cal F}_{\mu \nu}$ is not gauge covariant. However, the gauge invariance of the curvature $R_5$ can kept in the following two cases:

{\bf Case 1:} One of the gauge field for example $a_{R \mu}(x)$ is abelian, which is valid for the electroweak interaction. In this case ${\cal F}_{\mu \nu}$ reduces to the gauge covariant form. 

\begin{eqnarray}
{\cal F}_{\mu \nu} &=& f_{L\mu\nu}  +f_{R\mu\nu} 
\end{eqnarray}

This case enables us to construct the Einstein-Electroweak-quark-lepton system in the next section, when we couple the generalized gravity to quark-leptons.

{\bf Case 2:}  If we choose the left handed gauge field to be proportional to the right one with a constant factor $a_{L\mu} = \alpha a_{R \mu} = \alpha a_{\mu}$, we obtain the following covariant field strength  
\begin{eqnarray}
{\cal F}_{\mu \nu} &=& (\alpha +1)(\partial_\mu a_\nu - \partial_\nu a_\mu)  + m (\alpha-1)^2[a_{\mu}, a_{\nu}]~~~~
\end{eqnarray}
In the next Section, we will show that Case 1 enables us to unify generalized  Einstein gravity and electroweak interactions of quarks and leptons. And in the Case 2, the gauge fields are the same on both sheets of space-time. The same gluons will couple to the left- and right- handed chiral quarks, thus unifying gravity and QCD.  We also note that the results of Viet-Du \cite{VietDu2015}  are confirmed with the new spectral triple used in the present paper. 

\section{Einstein-Yang-Mills-Dirac systems}

In this section, we specialize to the Einstein-Yang-Mills-Dirac systems emerging from the generalized gravity coupled to the chiral spinors of quarks and leptons. The gauge fields contained in the generalized gravity, will now be non-Abelian and will enable us to incorporate weak and strong interactions of the standard model with appropriate colors and flavors. The couplings of gravity, gauge fields and Brans-Dickey scalar coupled to chiral spinors emerge naturally with some new interactions in contrast to our previous papers \cite{VietDu2015, QuyNhon2015, Viet2015} where these couplings were introduced as arbitrary and independent.  
The required total action $ S_{tot}$ in our framework is derived from the sum of the actions generalized from gravity sector $S_{HE}$ and the gravity-chiral spinor one $S_\psi$,

\begin{equation}
S_{tot} = S_{HE}(5) + S_\Psi
\end{equation}

\subsection{Generalized gravity coupled to the chiral spinors}

To obtain the generalized Lagrangian that contains  gravity coupled to chiral spinors, we begin with Eq.(\ref{eq:R5}) and the generalized covariant Dirac operator ${\cal D}$, where
\begin{equation} \label{GenCovDirac}
 {\cal D} = D + \Omega = D - {1 \over 8} \Gamma^C \Omega_{ABC} [\Gamma^A, \Gamma^B]. 
 \end{equation}
 
Then the generalized Dirac Lagrangian and its corresponding  action are given by
\begin{widetext}
\begin{eqnarray} \label{PsiL}
{\cal L}_\Psi&=& {\bar \Psi} (i {\cal D}+M) \Psi + h.c = {\bar \Psi} (iD +M)  \Psi +{\bar \Psi} \Omega \Psi + h.c= {\cal L}_D +{\cal L}_m + {\cal L}_\Omega(1) +  {\cal L}_\Omega(2),\\
S_{\Psi} &=& \int dx^4 \sqrt{-det g} \exp{({ \sqrt{2}\sigma(x) \over M_{Pl}(5)}+ \kappa_0)} {\cal L}_\Psi.
\end{eqnarray}
\end{widetext}
where $M$ is an arbitrary mass parameter to be chosen later and 
\begin{subequations}
\begin{eqnarray}\label{LDM}
&{\cal L}_D  =&  \sum_{I=L,R}  {\bar \psi}_I \gamma^a e^\mu_a (i\partial_\mu +  m a_{I\mu}(x) )\psi_I + h.c.~~~~~~~ \\
& {\cal L}_m  = &
  {\bar \psi}(m \exp{(-{ \sqrt{2}\sigma(x) \over M_{Pl}(4)}- \kappa_0)}+M) \psi + h.c. \\
& {\cal L}_\Omega(1) &= - {1 \over 8}{\bar  \psi}\gamma^c \omega_{abc} [\gamma^a, \gamma^b] \psi + h.c. \\
& {\cal L}_\Omega(2) &= {\sqrt{2} \over  M_{Pl}}{\bar \psi} \gamma^a e_a^\mu \partial_\mu \sigma(x) \psi +.h.c.
\end{eqnarray}
\end{subequations}
If $\kappa_0 \not = 0$, the factors of the kinetic and spinor connection terms in the actions $S_D$ and $S_\Omega(1)$ are not the usual ones. 

Nevertheless, it is possible to absorb those factors by redefining the chiral spinors
\begin{equation}\label{eq:spinredef}
\Psi \rightarrow e^{- {\kappa_0 \over 2}} \Psi
\end{equation}

So after the redefinition of the chiral spinors in Eq.(\ref{eq:spinredef}), retaining only the first order in $M^{-1}_{Pl}(4)$, we obtain the following results for actions from the Lagrangians in Eqs.(\ref{PsiL}) and (\ref{LDM})
\begin{widetext}
\begin{subequations}
\begin{eqnarray}
S_\Psi &=& S_\psi + S_{\sigma} + S_g + S_f \label{eq:SPsi}\\
S_\psi  & = & \int dx^4 \sqrt{-det g}~~  {\bar \psi} (i\gamma^a e^\mu_a \partial_\mu +  m e^{-\kappa_0} +M - {1 \over 8}\gamma^c \omega_{abc} [\gamma^a, \gamma^b]   )\psi + h.c. \label{eq:Spsi}\\
S_\sigma&=& {\sqrt{2} \over M_{Pl}} \int dx^4 \sqrt{-det g}~~{\bar \psi} [\gamma^a e_a^\mu (\partial_\mu \sigma(x) + \sigma(x) \partial_\mu) - {1 \over 8} \gamma^c \omega_{abc} [\gamma^a, \gamma^b] \sigma(x) ] \psi + h.c.\label{eq:Ssigma}~~~~ \\
S_g &=& m \int dx^4 \sqrt{-det g} (1+ {\sqrt{2}\sigma(x) \over M_{Pl}(4)})  \sum_{I=L,R}  {\bar \psi}_I \gamma^a e^\mu_a  a_{I\mu}(x) \psi_I + h.c. \label{eq:Sg} \\
S_f &=& {1 \over 16} \int dx^4 \sqrt{-det g}~ \bar \psi (f_{L\mu \nu}+ f_{R\mu \nu})\sigma^{\mu \nu} \psi + h.c., \label{eq:Sf}\end{eqnarray}
\end{subequations}
\end{widetext}
where $ \sigma^{\mu \nu} = i [\gamma^\mu, \gamma^\nu].$

Let us discuss the physical meaning of the above terms. 
$S_\psi$ describes the chiral spinors
in the curved 4D space-time with a mass of $m e^{-\kappa_0}+ M$. Since the quark-lepton masses are generated by the Higgs mechanism, we will choose $M= - m e^{-\kappa_0}$ to keep the quark-lepton to be massless as in the Standard Model.

$S_\sigma$ is responsible for quark-lepton pair creation of the Brans-Dicke scalar. The effect is difficult to observed experimentally, as the coupling constant is proportional to the inverse of the Plank mass.
 
$S_g$  contains gauge fields and its discussion will follow in the next subsection B. The action $S_f$ is a gauge invariant giving a significant contribution to the magnetic momentum of quark-leptons. It is not required by the first principle in the minimal coupled Yang-Mills-Dirac systems. In our framework, it is inherent. The discussion about its contributions to the magnetic momentum of quark-leptons will be discussed elsewhere.

\subsection{The Einstein-Electroweak-Quark-Lepton system}

We note that the gauge symmetry group for weak interactions in Yang-Mills approach is $SU(2) \times U(1)$ requiring  a non-Abelian  $SU(2)$ and an abelian $U(1)$ gauge fields. This leads us to Case 1 described in Section III in the framework the generalized gravity.  Focusing on quark and lepton families including the right-handed neutrino, we have the following left- and right-handed chiral quark-lepton representation,

\begin{eqnarray}
\psi_{LA} & =& \begin{bmatrix}
q^c_L \cr
l_L \cr
\end{bmatrix}_A ~~,~~
\psi_{RA} = \begin{bmatrix}
u^c_R \cr
d^c_R \cr
e_R \cr
\nu_R \cr
\end{bmatrix}_A,
\end{eqnarray}
where $c=1,2,3$ is the color index, $A$ is the family index, the number of which we leave arbitrary.

First, we introduce the usual physical gauge fields $W_\mu(x)$ and $B_\mu(x)$ as follows
\begin{eqnarray}\label{physicala}
a_{L\mu} &=& {1 \over M_L} ( W^a_\mu(x) {{\bf \tau^a} \over 2} \otimes {\bf 1_4}  - {M_L \over M_R} {Y_L \over 2} B_\mu(x) \otimes {\bf 1_2})  \otimes {\bf 1}_{N_F} , \nonumber \\
a_{R\mu} &=& - {1 \over M_R}  {Y_R \over 2} B_\mu(x)  \otimes {\bf 1}_{N_F},
\end{eqnarray}
where $Y_{L,R}$ are the hypercharge operators acting on the left and right-handed chiral quark-leptons. Since the gauge fields $a_{L,R \mu}$ are operators acting on the chiral quark-leptons, we must have include the corresponding unit matrices in Eqs.(\ref{physicala}). 

Now we choose the rescaling mass parameters for the gauge fields 
\begin{equation}
M_L =  {m \over g}  ~~,~~ M_R= {m \over g'}
\end{equation}

Substituting (\ref{physicala}) into Eq.(\ref{eq:Sg}) we obtain
\begin{widetext}
\begin{eqnarray}\label{LDM2}
S_g  &=&\int dx^4 \sqrt{-det g}~ (g'  {\bar \psi} \gamma^\mu  {Y \over 2} B_\mu(x) \psi -g {\bar \psi}_L \gamma^\mu  W^a_\mu(x) {{\bf \tau}^a \over 2} \psi_L)  \nonumber \\
&& +{\sqrt{2}\over M_{Pl}(4)} \int dx^4 \sqrt{-det g}~  \sigma(x) (g'  {\bar \psi} \gamma^\mu  {Y \over 2} B_\mu(x) \psi -g {\bar \psi}_L \gamma^\mu  W^a_\mu(x) {{\bf \tau}^a \over 2 } \psi_L), 
\end{eqnarray}
\end{widetext}
where the sum is taken over the color, flavor and family indexes. We note that the first term contains the usual electroweak couplings of the gauge fields to chiral spinors in the Standard Model. The second contains interaction terms of the gauge, quarks and leptons with the Brans-Dicke scalar $\sigma(x)$.

The hyper charge operators are the same for all the families and hence we specify them to be
\begin{eqnarray}
Y_L &=& \begin{bmatrix}
{1 \over 3} {\bf 1}_3 & 0 \cr
0 & -1 \cr
\end{bmatrix},\nonumber \\
Y_R &=& \begin{bmatrix}
{4 \over 3} {\bf 1}_3 & 0 & 0 &0 \cr
0 & -{2 \over 3} {\bf 1}_3 & 0 & 0\cr 
0 & 0 & -2 & 0\cr
0& 0 & 0 & 0 \cr
\end{bmatrix} ,
\end{eqnarray}

From Eqs.(\ref{physicala})- (\ref{TrR5}) and retaining only the first order of $\sigma(x)$ we obtain the generalized Hilbert-Einstein action for this case
\begin{widetext}
\begin{eqnarray} \label{SHESM}
S_{HE}(5) &=&
\int dx^4 \sqrt{-det g} ((M^2_{Pl}+ \sqrt{2} M_{Pl} \sigma(x)) r_4 -  { (M^2_{Pl} + 3\sqrt{2} M_{Pl} \sigma(x))\phi^2_0 \over 128m^2} ( g^2 Tr(W_{\mu \nu} W^{\mu\nu}) \nonumber \\
&& - {g'^2 \over 4} \sum (2Y^2_L + Y^2_R) B_{\mu \nu} B^{\mu\nu})), 
\end{eqnarray}
where the sum is taken over all the hyper charges of chiral quark-leptons and averaged on the families. Furthermore, in order to have the correct kinetic terms for the gauge fields, we find the following relations between the parameters
\begin{equation} \label{eq:g1g2}
g = {4\sqrt{2}m \over  M_{Pl} \phi_0 }  ~,~~ g= {g' \sqrt{\sum(2Y^2_L + Y^2_R)} \over 2} = g'\sqrt{10/3}, 
\end{equation}
where the factor $2$ in front of $Y^2_L$ is due to the fact that the left-handed quark-leptons always appear in doublet.

These relations lead to the following simplified form of Eq.(\ref{SHESM})
\begin{eqnarray}
S_{HE}(5)&=&
\int dx^4 \sqrt{-det g}~[ M^2_{Pl}  r_4 -
    {1 \over 4}( Tr(W_{\mu \nu} W^{\mu\nu})
   +  B_{\mu \nu} B^{\mu\nu}) \nonumber \\
 && +
   \sqrt{2} \sigma(x) (M_{Pl}(4)r_4 - {3 \over 4 M_{Pl}(4)} (Tr(W_{\mu \nu} W^{\mu\nu})
      +  B_{\mu \nu} B^{\mu\nu})) 
   ]   
\end{eqnarray}
\end{widetext}

In the above equation, the first term is the usual kinetic term for gravity and electroweak gauge fields. The second term represents interactions of the Brans-Dicke field $\sigma(x)$ with gravity, $W_\mu(x)$ and $B_\mu(x)$ fields. It is remarkable that the gauge fields are part of the generalized gravity. 
So, the full action of the Einstein-Electroweak-quark-lepton system can be derived from the generalized gravity and its coupling to the chiral quark-leptons.

Let us determine the energy scale, where the above coupling constant relations  become valid.
From Eqs.(\ref{eq:g1g2}) we have the prediction
\begin{equation}
\sin^2{\theta_W} = {g'^2 \over g^2 + g'^2} = {3 \over 13} = 0.23077,
\end{equation}
which has been given in \cite{Bes2004} from different considerations. 

Since this value is pretty close to the experimental value $\sin^2{\theta_W} = 0.231$, we can expect that the model is applicable for the Standard Model at currently accessible energy. It is remarkable is the gauge fields are derived from the generalized gravity.  
In this model, Higgs field must be added manually or generated dynamically.

\subsection{The Einstein-QCD system}
The Einstein-QCD system has the gauge group $SU(3)_{QCD}$ acting in the same way with the left and right-handed quarks.  This leads us to Case 2 discussed in Section III  with gauge fields as part of the generalized  Hilbert-Einstein action.

The quark sector of this system is represented by the following chiral spinors
\begin{eqnarray}
\psi_L &=& \begin{bmatrix}
u^c_L \cr
d^c_L \cr
\end{bmatrix} ~~,~~ \psi_R =\begin{bmatrix}
u^c_R \cr
d^c_R \cr
\end{bmatrix}.
\end{eqnarray} 

Next we introduce the physical gauge  fields representing color 
\begin{eqnarray}\label{physicalc}
a_{L\mu} &=&  \alpha a_{R\mu} = {\alpha \over m_{c}} C_\mu (x) \otimes {\bf 1}_6,
\end{eqnarray}
where $m_c$ is a mass parameter and $C_\mu(x) = C^a_\mu(x) {\lambda^a \over 2}$ is the color gauge field. The Gell Mann matrices $\lambda^a, a=1,..8$ are used as generators of the gauge group $SU(3)_{QCD}$.

Using Eqs.(\ref{physicalc}) and Eq.(\ref{eq:Sg}) we obtain the following action
\begin{widetext}
\begin{eqnarray}\label{eq:SgQCD}
S_g &=&  g_S \int dx^4 \sqrt{- det g}~   [{\bar \psi} \gamma^a e^\mu_a(x)  C_\mu(x)  \psi + {\sqrt{2} \sigma(x) \over M_{Pl}(4)}{\bar \psi} \gamma^a e^\mu_a(x)  C_\mu(x)  \psi \nonumber \\
&&
+ (\alpha -1) ({\bar \psi}_L \gamma^a e^\mu_a(x) C_\mu(x) \psi_L  +
 {\sqrt{2} \sigma(x) \over M_{Pl}(4)}{\bar \psi}_L \gamma^a e^\mu_a(x)   C_\mu(x) \psi_L )],
\end{eqnarray}
\end{widetext}
where the mass parameter $m_c$ is defined to be
\begin{equation}
m_c= {m \over g_S}.
\end{equation}

The first term is the coupling of between gluon gauge field and chiral quark in presence of gravity. The second term contains quark-gluon interactions with the Brans-Dicke scalar in the curved space-time. The third term violates P-conservationwhen $\alpha \not=1$.

By retaining only the first order of $\sigma(x)$, we can explicitly calculate the
generalized Hilbert-Einstein action as follows 
\begin{widetext}
\begin{eqnarray}\label{QCDaction1}
S_{HE}(5) &=& \int dx^4 \sqrt{-det g}~(1+ {\sqrt{2} \sigma(x) \over M_{Pl}(4)}) M^2_{Pl}(4)  r_4 
\nonumber \\
&& - {M^2_{Pl} \over 4} (1+ 3 {\sqrt{2} \sigma(x) \over M_{Pl}(x)}) g^{\mu \rho} g^{\nu \tau} (\alpha+1)^2 {g_S^2 \phi_0^2 \over  m^2}
 Tr (C_{\mu \nu} C_{\rho \tau}) \nonumber \\  C_{\mu \nu} &=& \partial_\mu C_\nu -\partial_\nu C_\mu+ g_S {(\alpha -1)^2 \over \alpha +1} [C_\mu, C_\nu].  
\end{eqnarray}
In order to have the correct factors of the kinetic term and to keep the gauge invariance, we must have the following relations
\begin{subequations}
\begin{eqnarray} 
&&{(\alpha -1)^2 \over \alpha +1} = 1 ~,~ \alpha = 3 \label{eq:QCDrel1} \\
&& g_S= {m \over  4 M_{Pl}(4) \phi_0} \label{eq:QCDrel2}
\end{eqnarray}
\end{subequations}
With the above conditions, the action and the field strength $C_{\mu \nu}$ in Eq. (\ref{QCDaction1}) takes the form

\begin{eqnarray}\label{QCDaction}
S_{HE}(5)&=& \int dx^4 \sqrt{-det g}~( M^2_{Pl} r^4 - {1 \over 4} Tr (C_{\mu \nu}C^{\mu \nu}) +  {\sqrt{2}\sigma(x) \over M_{Pl}(4)} ( M^2_{Pl}(4) r_4 - { 3 \over 4}  Tr C_{\mu \nu}C^{\mu \nu} ),\nonumber \\
C_{\mu \nu} &=& \partial_\mu C_\nu -\partial_\nu C_\mu+ g_S  [C_\mu, C_\nu].
\end{eqnarray}
\end{widetext}
The first term is the usual action of gluon fields coupled to gravity. 
The second term represents interactions of the gluon field and the  Brans-Dicke scalar in the curved space-time.

Since in Eq(\ref{eq:QCDrel2}), the strong coupling constant depends on two unknown parameters $m$ and $\phi_0$, we do not have a prediction for the coupling constant as in the Einstein-electroweak-quark-lepton model. We can use these two parameters to tune the model to be applicable at the currently accessible energy.

The new feature of the Einstein-QCD system constructed from the generalized gravity is the intrinsic P-violation. In fact, if $\alpha =1$, the commutator in the field strength $C_{\mu \nu}$ vanishes, thus the gauge invariance can not be kept. So, the parity violation is inherent in the constructed system.

Since the currently available data suggests that the strong interaction does not violate parity at least up to the energy scale of $10^2-10^3 GeV$, we can choose to transfer the P-violation in Eq.(\ref{eq:SgQCD}) into the gravity sector. To do so, we must go back to our general theory with bigravity of right-handed and left-handed vierbeins in Eqs.(\ref{eq:SHE}) and (\ref{eq:R5}). If instead of the condition (\ref{eq:samegravity}) we assume the following
\begin{equation}
e^\mu_{L_a}(x) = {1 \over 3} e^\mu_{Ra}(x) = {1 \over 3} e^\mu_a(x),
\end{equation}
to lead to the following parity  non-violation action
\begin{widetext}
\begin{eqnarray}\label{eq:SgQCD2}
S_g &=&  g_S \int dx^4 \sqrt{- det g}~   [{\bar \psi} \gamma^a e^\mu_a(x)  C_\mu(x)  \psi + {\sqrt{2} \sigma(x) \over M_{Pl}(4)}{\bar \psi} \gamma^a e^\mu_a(x)  C_\mu(x)  \psi
\end{eqnarray}
\end{widetext}
Then the P-violation will appear in the action $S_\psi$.
Although, this possibility might lead to new physical consequences, it does not contradict to the particle physics under the $TeV$ energy range. 

\section{Conclusions}

In this article, we have constructed the coupling of chiral spinors and generalized gravity within the framework of discretized Kaluza-Klein theory with one internal extra dimension having only two points. This is a different approach than the traditional Kaluza-Klein theory to unify gravity with the gauge fields. The discrete dimension can describe two copies of space-time of the left and right-handed chiral fermions proposed by Connes and Lott \cite{CoLo} or other discrete properties of quark-lepton as in Refs.\cite{QuyNhon2015, Viet2015}.  Starting with the Dirac operator in Eq.(\ref{DIRAC2}) and the not fully anti-symmetric wedge product, we have derived a general model containing a finite gravity sector with bigravity, two possibly nonabelian gauge fields and one Brans-Dicke scalar. We also recovered the same result of Viet-Du in this framework, at least one gauge field must be abelian or both gauge fields must be the same.

Thus, we are able to construct different realistic Einstein-Yang-Mills-Dirac systems out from generalized gravity coupled to chiral spinors. All the gauge couplings and kinetic terms of quark-leptons can be derived from the generalized gravity. In addition, a non-Higgs mass term together with a magnetic momentum interactions of quark-leptons is inherent in the framework. While the quark-lepton masses are zero and the value of $\phi_0$ is large, the contribution of the magnetic momentum term is significant.

The Einstein-electroweak-quark-lepton system, as one special case of Viet-Du's theorem, leads to a value $\sin^2{\theta_W} = 0.23077$ very close to the experimental one $0.231$ indicating that the model is applicable to the realistic particle physics. At the same time, with the experimental value of the coupling constant $g$ at the electroweak energy scale, our model implies a large value of $m$.

The Einstein-QCD system as another case of Viet-Du's theorem, also leads to a realistic model with a new parity violation interaction, which can be transferred to the gravity sector by allowing the gravitation interacts differently with left- and right-handed chiral fermions. The parity violation coupled to gravity may be of relevance to the long standing problem of baryon-anti Baryon asymmetry.

Since the condition of Viet-Du's theorem can not hold at the same time with one single discrete extra dimension, one can consider both strong, electroweak interactions and Higgs as components of gravity in the discretized Kaluza-Klein theory with two discrete extra dimensions as proposed by one of us \cite{QuyNhon2015, Viet2015}. In such an extended model, the coupling constants and Higgs mass will be related.

Thanks are due to Pham Tien Du and Do Van Thanh (Department of Physics, Hanoi University of Science, VNU) for their participation in the research group. The supports of ITI, VNU and Department of Physics, Hanoi University of Science, VNU are greatly appreciated. The work of NSH is supported by NAFOSTED under grant No 103.03-2012.02.

\end{document}